\documentstyle [12pt] {article}

\newcommand{\beq}{\begin{equation}}
\newcommand{\eeq}{\end{equation}}
\newcommand{\beqs}{\begin{eqnarray}}
\newcommand{\eeqs}{\end{eqnarray}}
\newcommand{\prl}{Phys. Rev. Lett.}
\newcommand{\prd}{Phys. Rev. D}
\newcommand{\npb}{Nucl. Phys. B}
\newcommand{\plb}{Phys. Lett. B}

\begin{document}

\baselineskip 7.5mm

\begin{flushright}
\begin{tabular}{l}
ITP-SB-93-37    \\
hep-ph/9307344  \\
July, 1993
\end{tabular}
\end{flushright}

\vspace{8mm}
\begin{center}
{\Large \bf  A New Model for Fermion Masses in }\\
\vspace{4mm}
{\Large \bf  Supersymmetric Grand Unified Theories } \\
\vspace{16mm}

Alexander Kusenko\footnote{email: sasha@max.physics.sunysb.edu}
and Robert Shrock\footnote{email: shrock@max.physics.sunysb.edu;
partially supported by NSF grant PHY-91-08054}

\vspace{6mm}
Institute for Theoretical Physics  \\
State University of New York       \\
Stony Brook, N. Y. 11794-3840  \\

\vspace{20mm}

{\bf Abstract}
\end{center}

We present a simple model for fermion mass matrices and quark mixing in the
context of supersymmetric grand unified theories and show its agreement
with experiment. Our model realizes the GUT mass relations $m_d=3m_e$,
$m_s= m_\mu/3$, $m_b=m_\tau$ in a new way and is easily consistent with values
of $m_t$ suggested by MSSM fits to LEP data.

\vspace{35mm}

\pagestyle{empty}
\newpage

\pagestyle{plain}
\pagenumbering{arabic}

   Understanding the values of fermion masses and quark mixing remains one of
the most important unsolved problems in particle physics.  The magnitude of the
challenge is shown by the fact that at present it is not definitely clear
which approach one should take to the related matter of electroweak symmetry
breaking (EWSB).  Intensive efforts continue in at least two quite different
directions: (i) supersymmetry, and, in particular, the minimal supersymmetric
standard model (MSSM), in which EWSB is perturbative~\cite{susy}; (ii)
dynamical (and nonperturbative) EWSB, e.g. slowly running
technicolor~\cite{wtc}, as well as others such as compositeness.
Here we shall adopt the first
approach, specifically the MSSM, which may provide the
simplest way to understand the precision electroweak measurements from
LEP~\cite{lep} while stabilizing the Higgs sector in the standard model.  A
further motivation for the MSSM is that it achieves unification of gauge
couplings at a single scale $m_G$, a prerequisite for a supersymmetric grand
unified theory (SGUT).

  In GUT's such as SU(5) or SO(10), by well-known restrictions on
(renormalizable) Higgs
couplings, the elements of the lepton Yukawa matrix can be made equal to $1$ or
$-3$ times corresponding elements of down quark Yukawa matrix.  In this
approach, one can get a simple relation between the resultant
(running) lepton and down quark masses, evaluated at the GUT scale. One
interesting proposal~\cite{gj} is that at the GUT scale
\beq
m_d = 3 m_e \ , \ \ m_s = \frac{m_\mu}{3} \ , \ \ m_b = m_\tau
\label{mrel}
\eeq
It is important to determine which (experimentally acceptable)
forms for fermion mass matrices yield this relation.  Since 1979, essentially
only one such form (up to trivial equivalence transformations) has been found.
With ${-\cal L}_m = v_u
\bar \psi_{u,L} Y_u \psi_{u,R} + v_{de}( \bar\psi_{d,L} Y_d \psi_{d,R} +
 \bar\psi_{e,L} Y_e \psi_{e,R}) + h.c.$, this form has GUT
scale Yukawa matrices
\beq
Y_u=  \left (\begin{array}{ccc}
               0 & A_u & 0 \\
               A_u & 0 & B_u \\
               0 & B_u & C_u \end{array} \right  )
\label{yuo}
\eeq
\smallskip
\beq
Y_d=    \left ( \begin{array}{ccc}
               0 &  A_d e^{i\phi} &   0 \\
               A_d e^{-i\phi} & B_d & 0 \\
               0 & 0 & C_d \end{array} \right )
\label{ydo}
\eeq
\smallskip
\beq
Y_e=  \left ( \begin{array}{ccc}
                0 & A_d &  0 \\
               A_d & -3 B_d & 0 \\
               0 & 0  & C_d \end{array} \right  )
\label{yeo}
\eeq
Interesting studies of SGUT Yukawa matrices, and this ansatz in particular,
have been carried out recently using renormalization group equations (RGE's)
in the MSSM~\cite{r,dhr,bbo} (where $v_u = 2^{-1/2}v \sin \beta$,
$v_{de} = 2^{-1/2}v \cos \beta$, with $v = 2^{-1/4}G_F^{-1/2} = 246$ GeV).
A general feature of this ansatz is that it tends to
require a rather large top quark mass, $m_t$.
Further recent works have considered more complicated mass matrices~\cite{giu}
and higher-dimension operators~\cite{dhr2} which do not, in
general, yield (\ref{mrel}) (see also Ref. \cite{rrr}).

   In this paper we exhibit a new and quite different model for fermion mass
matrices which yields the GJ mass relation (\ref{mrel}).
We show that this model agrees with
experiment, in particular with values of $m_t$ around 135 GeV, in
the range suggested by MSSM fits to the LEP data~\cite{lep,mt}.  Our model is
defined at the SGUT scale by
\beq
Y_u=  \left (\begin{array}{ccc}
                  0 & A_u & 0 \\
                  A_u & B_u & 0 \\
                  0 & 0 & C_u \end{array}   \right  )
\label{yu}
\eeq
\smallskip
\beq
Y_d=     \left ( \begin{array}{ccc}
                  0 &  A_d e^{i\phi} &  0 \\
                  A_d e^{-i\phi} & B_d & B_d \\
                  0 & B_d & C_d \end{array}  \right )
\label{yd}
\eeq
\smallskip
\beq
Y_e=  \left ( \begin{array}{ccc}
                   0 & A_d & 0 \\
                   A_d & -3 B_d & -3 B_d \\
                   0 &-3 B_d & C_d \end{array} \right  )
\label{ye}
\eeq
A natural setting for this ansatz would be an SO(10) SGUT, where one can obtain
symmetric (complex) Yukawa matrices by using {\bf 10} and {\bf 126}
Higgs representations to couple to the fermion ${\bf 16} \times {\bf 16}$
bilinear.  If one had a
fundamental theory of fermion masses, one would presumably
be able to derive the forms of the $Y_f$ and also the values
of the parameters from first principles.  In the
absence of such a theory, we believe that models such as ours can give valuable
hints about the underlying physics~\cite{sym}.  Our mass matrix model
has seven real parameters $A_u$, $B_u$, $C_u$, $A_d$, $B_d$, $C_d$,
and $\phi$ (plus $\tan \beta$) to describe the
nine fermion masses in the $u,d,e$ sectors and the four angles parametrizing
the quark mixing matrix $V$ (plus $\tan \beta$); hence it makes six
predictions. (Because of the uncertainties regarding neutrino masses and
mixing, we do not consider these here.)

    We first show that this ansatz yields the GUT scale GJ mass relation
(\ref{mrel}).  The $Y_f$, $f=u,d,e$
are diagonalized by unitary transformations $U_f$: $U_f Y_f U_f^\dagger =
diag\{\lambda_{f,1},\lambda_{f,2},\lambda_{f,3}\}$, where $\lambda_{f,j}=\pm
m_{f,j}/v_f$ (with the quantities defined at $\mu=m_G$, and $v_d=v_e \equiv
v_{de}$).  Without loss of generality, one may pick $\lambda_{f,3} > 0$.
For $Y_d$ one can take either of the choices
$(\lambda_{d,1},\lambda_{d,2},\lambda_{d,3})=(+,-,+), \ (-,+,+)$;
correspondingly, $(\lambda_{e,1},\lambda_{e,2},\lambda_{e,3})=(-,+,+), \
(+,-,+)$.  To leading order in small fermion mass ratios, we find that $C_d =
C_e$ $\Rightarrow m_b = m_\tau$.  Next, $B_d = \lambda_{d,2} = \mp
m_{s}/v_{de}$ and $\lambda_{e,2} = -3\lambda_{d,2}$, which together imply
$m_s = m_\mu/3$. Finally, $A_d = \sqrt{-\lambda_{d,1}\lambda_{d,2}}=
\sqrt{m_d m_s}/v_{de}$ and $(Y_e)_{12} = (Y_e)_{21} =
\sqrt{-\lambda_{e,1}\lambda_{e,2}}=\sqrt{m_e m_\mu}/v_{de}$  so that the
property $(Y_e)_{12}=A_d$ implies $\sqrt{m_e m_\mu}=\sqrt{m_d m_s}$ and hence
$m_d = 3 m_e$.  (These relations hold for either
set of sign choices.) This proves our assertion. We will use the choice
$(-,+,+)$ for $Y_d$ since it gives a slightly better fit to the data on quark
mixing.  For $Y_u$ one may also take $(\lambda_{u,1},\lambda_{u,2},
\lambda_{u,3})=(+,-,+)$ or $(-,+,+)$, and one finds
$C_u=\lambda_{u,3}=m_t/v_u$, $B_u = \lambda_{u,2}=\mp m_c/v_u$, and
$A_u=\sqrt{-\lambda_{u,1}\lambda_{u,2}}=\sqrt{m_um_c}/v_u$. The different sign
choices for the $\lambda_{u,j}$ will imply picking $\phi$ values in the fit
which differ by $\pi$; for definiteness, we use $(+,-,+)$ for the
$\lambda_{u,j}$.

    To test the model, we evolve the Yukawa matrices down from the SGUT scale
to the electroweak scale, $m_{_{EW}}$ and diagonalize them there.  We shall
take $m_{_{EW}}=m_t$, but our results are not very sensitive to this choice.
We use the illustrative value $m_t(m_t) = 135$ GeV (i.e., pole mass
$M_t=m_t(m_t)[1+4\alpha_3(M_t)/(3\pi) + O(\alpha_3(M_t)^2] \simeq 142$ GeV),
consistent with MSSM fits to LEP data ~\cite{lep,mt}. To avoid
complications involving further model-dependence, we follow a common
simplification~\cite{dhr,bbo} of approximating the full SUSY mass spectrum by
a single mass scale, $m_{SUSY}$ \cite{fsusy} with $m_{SUSY}= m_t$.
Given the theoretical
uncertainties associated with the SUSY spectrum and threshold corrections,
1-loop RGE's are sufficiently accurate in the range from
$m_G$ to $m_{SUSY}$.  The inputs $\alpha_{em}(m_Z)^{-1}=127.9 \pm 0.2$
and $\sin^2 \hat\theta_W = 0.2325 \pm 0.008$ \cite{pdg92},
which give $\alpha_1(m_Z)^{-1}=58.90 \pm 0.11$ and
$\alpha_2(m_Z)^{-1}=29.74 \pm 0.11$, together with the assumptions that the
lightest Higgs has $m_h \simeq m_Z$ and $m_{SUSY} \simeq m_t = 135$ GeV
lead to 1-loop unification in the MSSM at
$m_G = 1.3 \times 10^{16}$, with $\alpha_G^{-1}=24.8$.
We take $\alpha_3(m_Z)=0.125$, consistent with
current measurements \cite{lep,beth} and with MSSM gauge coupling
unification.  (Using the 3-loop QCD RGE's~\cite{tar}, this gives
$\Lambda^{(5)}_{\overline{MS}}=302$ MeV and $\alpha_3(m_t)=0.118$.)

    An interesting property of the model which we have established by analytic
and numerical solutions of the RGE's is that although the
values of the parameters change, the forms of $Y_u$ and $Y_e$ are preserved
to high accuracy under evolution from $m_G$ to $m_{_{EW}}$.  For $Y_d$, the
relation $(Y_d)_{21}=(Y_d)_{12}^*$ and the equality $(Y_d)_{22}=(Y_d)_{23}$ are
maintained quite accurately, while $|(Y_d)_{32}|$ decreases (by about 35 \%)
relative to $|(Y_d)_{23}|$.
Keeping dominant terms (and with $\dot f(t) \equiv (16
\pi^2)^{-1} d f/dt$, $t = \ln (\mu/m_G)$), the RGE's for the Yukawa matrices
thus reduce to RGE's for the nonzero elements, which are
\beq
\dot X_u=X_u (- \sum_j c_{u,j} g_j^2)
\label{aubudot}
\eeq
for $X_u=A_u, \ B_u$;
\beq
\dot C_u=C_u (- \sum_j c_{u,j} g_j^2+ 6 C_u^2)
\label{cudot}
\eeq
\beq
\dot X_d=X_d(-\sum_j c_{d,j} g_j^2 )
\label{adbddot}
\eeq
for $X_d=A_d, \ (Y_d)_{22}, \ (Y_d)_{23}$;
\beq
\dot Z_d=Z_d(-\sum_j c_{d,j}g_j^2 +  C_u^2)
\label{cddot}
\eeq
for $Z_d = (Y_d)_{32}, \ (Y_d)_{33}$; and
\beq
\dot X_e = X_e (-\sum_j c_{e,j} g_j^2)
\label{xedot}
\eeq
for $X_e=A_e, \ B_e, \ C_e$, where for the MSSM in the relevant energy range
$\mu > m_{SUSY}$, $(b_1,b_2,b_3)=(\frac{33}{5},1,-3)$, \ \
$(c_{u,1},c_{u,2},c_{u,3})=(\frac{13}{15},3,\frac{16}{3})$, \ \
$(c_{d,1},c_{d,2},c_{d,3})=(\frac{7}{15},3,\frac{16}{3})$, \ \
$(c_{e,1},c_{e,2},c_{e,3})=(\frac{9}{5},3,0)$ \cite{con}.
To this order, the running of $\phi$ is negligibly small.

Manipulating these RGE's, we obtain the solutions
\beq
\frac{m_b(m_{_{EW}})}{m_\tau(m_{_{EW}})}=\frac{C_d(t_{_{EW}})}
{C_e(t_{_{EW}})}=\biggl (\frac{C_u(t_{_{EW}})}{C_u(t_G)} \biggr )^{1/6}
\rho_{_C}(t_{_{EW}})
\label{mbmtau}
\eeq
where
\beq
\rho_{_C}(t) = \prod_{j=1}^3 \biggl (\frac{\alpha_G}{\alpha_j(t)} \biggr )^
{\frac{c_{d,j}-c_{e,j}-(c_{u,j}/6)}{2 b_j}}
\label{p2}
\eeq
and $t_{_{EW}}=\ln(m_t/m_G)$.  We calculate $\rho_{_C}(t_{_{EW}})=1.92$.
Further, we define $\eta_f = m_f(m_f)/m_f(m_{_{EW}})$ for leptons and heavy
quarks, and $\eta_{uds}=m_f(1 \ GeV)/m_f(m_{_{EW}})$ for $u,d,s$ and take the
value $m_b(m_b)=4.19$ GeV, consistent with Ref. \cite{gl}.  Using 1-loop
QED and 3-loop QCD RGE's, we obtain $\eta_\tau=1.02$ and $\eta_b=1.55$, whence
$m_b(m_{_{EW}})/m_\tau(m_{_{EW}})=1.54$.  We successfully fit this ratio by
choosing an appropriate value of $C_u(t_G)$.  This is a nontrivial
test of the ansatz, since it is not, {\it a priori}, guaranteed that a
perturbatively acceptable value of $C_u(t_G)$ would enable one to fit the
observed $m_b/m_\tau$ mass ratio.  We find $C_u(t_{_{EW}})/C_u(t_G)=0.269$
from (\ref{mbmtau}); substituting this into
the solution to the RGE for $C_u$ \cite{f2}, we get $C_u(t_G)=4.48$.  (This is
consistent with the validity of perturbation theory since loop corrections are
small: $C_{u}^2/(16 \pi^2) = 0.13$.)  We then get $C_u(t_{_{EW}})=1.21$ and,
from the condition $m_t(m_t)=C_u(t_{_{EW}}) v \sin(\beta)/\sqrt{2}$, determine
$\tan \beta = 0.84$.

   From the RGE's, we obtain the solutions
\beq
\frac{A_d(t)}{A_e(t)}=\frac{B_d(t)}{B_e(t)}=\rho_{AB}(t)
\label{abeq}
\eeq
where $(Y_d)_{22}(t)=(Y_d)_{23} \equiv B_d(t)$ and
\beq
\rho_{AB}(t) = \prod_{j=1}^3 \biggl (\frac{\alpha_G}{\alpha_j(t)}
\biggr )^{\frac{c_{d,j}-c_{e,j}}{2 b_j}}
\label{rhoab}
\eeq
We calculate $\rho_{AB}(t=t_{_{EW}})=2.38$.  Now
\beq
\frac{m_s(m_{_{EW}})}{m_\mu(m_{_{EW}})}=
\frac{B_d(t_{_{EW}})}{ 3 B_e(t_{_{EW}})}=0.792
\eeq
Using $m_\mu(m_{_{EW}})=102.8$ MeV, the model predicts $m_s(m_{_{EW}})=81.4$
MeV, i.e., with our 3-loop $\eta_{uds}=2.94$, $m_s(1 \ GeV) = 239$ MeV.
This is consistent with the (upper range of the) determination
$m_s(1 \ GeV)=175 \pm 55$ MeV \cite{gl}.  Further we find
\beq
\frac{(Y_d)_{32}(t)}{(Y_d)_{23}(t)}=\biggl (\frac{C_u(t)}{C_u(t_G)}
\biggr )^{\frac{1}{6}}\prod_{j=1}^3 \biggl ( \frac{\alpha_G}{\alpha_j(t)}
\biggr )^{-\frac{c_{u,j}}{12 b_j}}
\label{y32y23}
\eeq

  We use the results $A_d(t)v_{de} = \sqrt{m_d(t) m_s(t)}$, $A_e(t)v_{de} =
\sqrt{m_e(t) m_\mu(t)}$ and eq. (\ref{abeq}) to find
$m_d(m_{_{EW}})/m_e(m_{_{EW}})=7.13$.  Using
$m_e(m_{_{EW}})=0.4876$ MeV, this yields $m_d(m_{_{EW}})=3.48$ MeV, and hence
$m_d(1 \ GeV)=10.2$ MeV, in agreement with the
value $m_d(1 \ GeV) = 8.9 \pm 2.6$ MeV \cite{gl}.

    Given that $Y_u$ retains its form to high accuracy under the evolution from
$m_G$ to $m_{_{EW}}$, the two remaining parameters in $Y_u$ are determined
via the relations $B_u(t_{ew})v_u=-m_c(t)$ and $A_u(t_{ew})v_u =
\sqrt{m_u(t) m_c(t)}$.  From $m_c(m_c)=1.27 \pm 0.05$ GeV \cite{gl}
we calculate $\eta_c= 2.32$, whence $m_c(m_{_{EW}})=0.547 \pm 0.022$ GeV.
With $m_u(1 \ GeV)= 5.1 \pm 1.5$ MeV we get $m_u(m_{_{EW}})=1.7 \pm 0.5$ MeV.
Diagonalizing at $m_{_{EW}}$, we have $v_uU_{u,L}Y_u U_{u,L}^\dagger =
M_u$, $v_d^2U_{d,L}Y_d Y_d^\dagger U_{d,R}^\dagger = M_d^2$, whence $V =
U_{u,L}U_{d,L}^\dagger$ for the Cabibbo-Kobayashi-Maskawa (CKM) matrix $V$
\cite{vrun}.  To leading order $|V_{us}| \simeq |\sqrt{m_u/m_c} -
e^{i\phi}\sqrt{m_d/m_s}|$ (with quantities evaluated at $\mu=m_{_{EW}}$
\cite{vrun}). We determine $\phi=75.5^\circ$ by fitting $|V_{us}|$.
Quoting as usual the rephasing-invariant $|V_{jk}|$ values, we get
\beq
\{ |V_{ij}| \} = \left ( \begin{array}{ccc}
                     0.9753 & 0.221 & 0.002 \\
                     0.221 & 0.9748 & 0.030 \\
                     0.006 & 0.030 & 0.9995 \end{array} \right )
\label{vfnum}
\eeq
in agreement with current data~\cite{pdg92,cleo2} (with $|V_{cb}|$ in the lower
range of acceptable values).
For the reparametrization-invariant CP violation parameter $J$ \cite{jarl}, we
get $J= 0.95 \times 10^{-5}$, again in agreement with data.
Further details will be given elsewhere~\cite{ks3}.


\begin{thebibliography}{99}
\bibitem{fsusy}{To do significantly better than the assumption of a single
$m_{SUSY}$  would require a
self-consistent evolution of the full theory down from the SGUT level, which
would predict both the supersymmetric particle mass spectrum and the usual
Yukawa mass matrices, while addressing issues such as the ``$\mu$ problem'',
doublet-triplet splitting, flavor-changing neutral operators, etc; inevitably,
this would involve more parameters and model-dependence.  The effect of using
$m_{SUSY}=1$ TeV as well as $m_{SUSY}=m_t=150$ GeV has been studied in
Ref. \cite{bbo}.}
\bibitem{f2}{Since $C_u >>$ all other terms in the $Y_f$, the dominant terms in
the 1-loop RGE for $C_u$ (\ref{cudot}) are the same for our model and that of
(\ref{yuo})-(\ref{yeo}); the solution is $(C_u(t)/C_u(t_G))^2 =
\rho(t)/[1+3(2\pi)^{-2}C_u^2(t_G)I(t)]$, where
$\rho(t)=\prod_{j=1}^3(\alpha_G/\alpha_j(t))^{c_{u,j}/b_j}$ and
$I(t)=\int_{t}^{t_G} \rho(t')dt'$.  For our values of $\alpha_3(m_Z)$ and
$m_{_{EW}}$, we get $\rho(t_{_{EW}})=13.2$ and $I(t_{_{EW}})=119.$}
\bibitem{vrun}{The running of $V$ from $m_{_{EW}}$ to the low energies where
its elements have been measured can be neglected.}
\end{thebibliography}
\end{document}